# 4π Combination of Gaussian Laser Beams

**Hocine.Djellout**

Laboratoire LPCQ, UMMTO, 15000 Tizi-Ouzou
e-mail : hocine.djellout@ummto.dz

tel : +213(0)558682673

***Résumé :***

*This work is a generalization of our previous study [1], which dealt with the coherent and incoherent combination of linearly polarized Gaussian laser beams operating in continuous-wave mode under a 2π focusing configuration. In the present work, the laser beams are focused within a solid angle that can reach 4π by using two opposing source planes, where all the lasers from both planes are focused onto the same point in the focal plane.*

*Our results show that the focal spot area obtained with either one or two source planes remains unchanged, despite the fact that, in the two-source-plane configuration, the numerical aperture (NA) of the system is twice that of the 2π focusing scheme. However, the intensity at the focal point is four times greater than that obtained with a single source plane. In addition, the longitudinal resolution is significantly improved, yielding a focal volume smaller than $(0.5\lambda)^3$ for a numerical aperture of NA = 0.895.*

*For incoherent beam combination, the focal volume remains practically the same in both the 2π and 4π focusing configurations, while the intensity at the focal point is twice that obtained in the 2π case. Furthermore, in the incoherent configuration, we observe a depth of focus reaching up to 30,000 λ for a system with a low numerical aperture.*

**Keywords:** *Incoherent beam combining (IBC), coherent beam combining (CBC), 4π focusing, high-power laser systems.*

## Introduction :

4π focusing is generally achieved either by means of a parabolic mirror **[2]** or by using two opposing high-numerical-aperture (NA) lenses **[3]**. It has found applications in various fields such as nanoscopy **[4]**, single-particle studies **[5]**, and others. A 4π configuration can also be realized by combining multiple laser beams, as implemented in large-scale facilities such as the LMJ and the NIF **[6,7]**, which are currently in operation for nuclear fusion research. To achieve extremely high laser intensities, 4π beam combination has also been proposed as a means of approaching the threshold for electron–positron ($e^-$–$e^+$) pair production from vacuum **[8]**.

The electromagnetic fields and intensity distribution in the focal region of non-paraxial beams are generally described using the Richards–Wolf formalism **[9]**. In this article, however, we employ a different model, namely the one previously developed in our earlier work **[1]**, which addressed the combination of laser beams over a solid angle smaller than 2π. The basic idea is straightforward: since the electric field and intensity of a Gaussian beam are known at every point in space, it is also possible to determine the field and intensity resulting from the combination of several laser beams. Indeed, the resultant field at any point in space is simply the sum of the fields contributed by all the laser beams at that point. Therefore, determining the total field and intensity becomes essentially a geometrical problem.

In this work, we present a theoretical study of the combination of multiple Gaussian laser beams in a 4π configuration using two opposing source planes. Each source plane contains several laser beams, all focused onto the same focal point. We first investigate the case of incoherent beam combining, in which the laser beams are not phase-locked. We then consider the case of coherent beam combining, where all the laser beams are mutually coherent.

## 1- Incoherent Beam Combining

The 4π configuration considered in this work is illustrated in Figure 1. It consists of two opposing source planes, each containing Gaussian laser beams with a minimum beam waist $w_0$. The coordinate system $(o_1, x_1, y_1, z_1)$ is associated with source plane 1, while $(o_2, x_2, y_2, z_2)$ is associated with source plane 2. The image plane is described by the coordinate system $(O, X, Y, Z)$. The center of each laser beam (i,j) in source plane 1 is located at coordinates $(x_i, y_j)$, whereas the center of laser beam (n,p) in source plane 2 is located at coordinates $(x_n, y_p)$. In order to maximize the combined intensity in the image plane, each laser beam is directed and focused toward the focal point (O). This is achieved using a lens with focal length $f_{ij}$ for the laser beams of source plane 1 and $f_{np}$ for those of source plane 2. These focal lengths are equal to the distance between the laser position in the corresponding source plane and the focal point (O), which is taken as the origin of the image-plane coordinate system. The distances between source planes 1 and 2 and the image plane are denoted by $h_{10}$ and $h_{20}$, respectively.

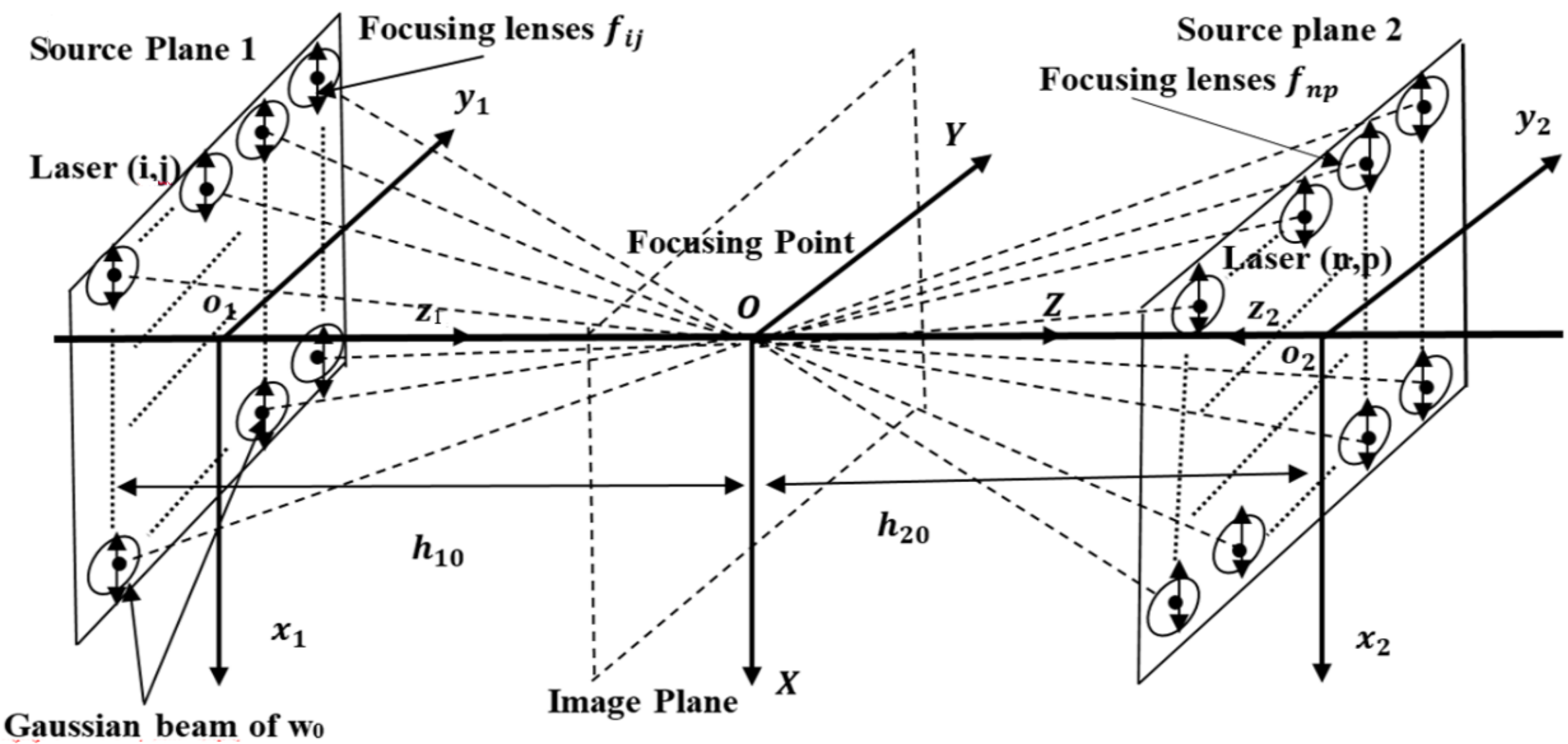

**Fig 1 : 4π configuration consisting of two opposing source planes.**

The electric field of a Gaussian beam with minimum waist $w_0$, centered at the origin $o_1$, is given by

$$\overrightarrow{E_{00}}(z_1, r) = \vec{\imath}\,\sqrt{I_0}\,\frac{w_0}{w(z_1)}\,exp\left[-\frac{r^2}{w^2(z_1)}\right]\,exp\left[i\left(kz_1 - arctan\left(\frac{z_1}{z_0}\right) + \frac{kr^2}{2R(z_1)}\right)\right] \quad \textbf{(1)}$$

$I_0$ is the on-axis intensity of the beam at $z_1 = 0$, $k = \frac{2\pi}{\lambda}$, $\lambda$ is the wavelength,

$$w(z_1) = w_0\sqrt{1 + \left(\frac{z_1}{z_0}\right)^2} \quad \textbf{(2)}$$

$w(z_1)$ denotes the beam radius at the position $z_1$, $z_0 = \frac{\pi\, w_0^2}{\lambda}$ represents the Rayleigh range,

$$R(z_1) = z_1\left[1 + \left(\frac{z_0}{z_1}\right)^2\right] \quad \textbf{(3)}$$

$R(z_1)$ is the radius of curvature of the Gaussian beam at the coordinate $z_1$, $r$ is the radial coordinate, and $arctan\left(\frac{z_1}{z_0}\right)$ is the Gouy phase shift. The intensity of the Gaussian beam is equal to $I_{00} = \left|\overrightarrow{(E_{00})}\ \overrightarrow{(E_{00})}^*\right| = I_0 \frac{{w_0}^2}{w(z_1)^2}\ exp\left[-2\frac{r^2}{w^2(z_1)}\right]$.

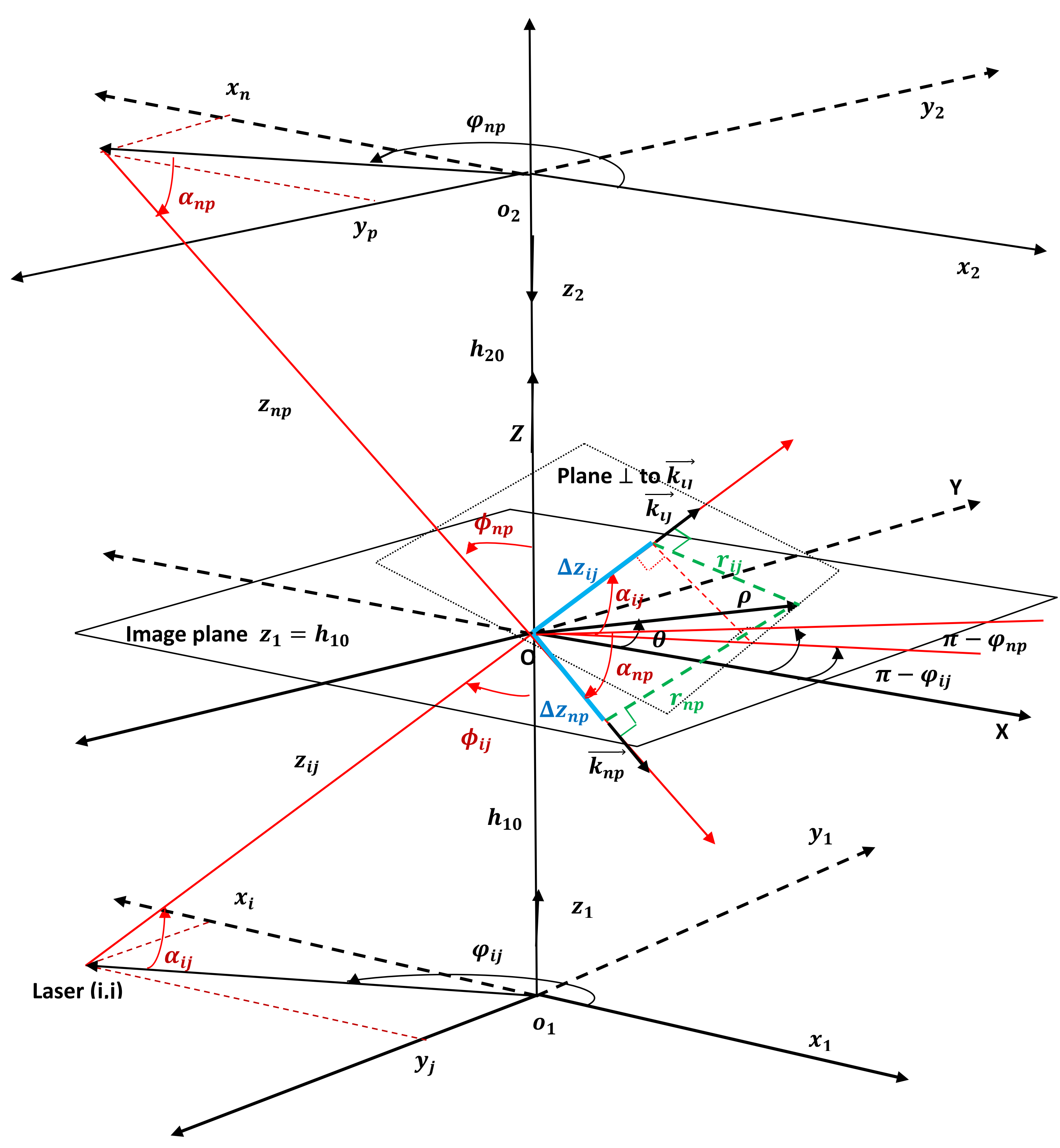

**Fig 2 : Geometrical determination of $\Delta z_{ij}$ , $\Delta z_{np}$ , $r_{ij}$ , $r_{np}$ at the plane $z_1 = h_{10}$**

To calculate the intensity of the (i,j) laser beam propagating from source plane 1 to the image plane (O,X,Y,Z), it is sufficient to determine $\Delta z_{ij}$ and $r_{ij}$, which were already derived in **[1]**. Similarly, to calculate the intensity of the (n,p) laser beam propagating from source plane 2 to the image plane, it is necessary to determine $\Delta z_{np}$ and $r_{np}$. Owing to the symmetry of the problem, as illustrated in Figure 2, the parameters $\Delta z_{ij}$ and $\Delta z_{np}$ have practically identical expressions. The same applies to $r_{ij}$ and $r_{np}$.

$$\Delta z_{ij} = \rho\ \cos(\theta - \varphi_{ij} + \pi)\ \cos(\alpha_{ij}) \qquad (4)$$

$$\Delta z_{np} = \rho\ \cos(\theta - \varphi_{np} + \pi)\ \cos(\alpha_{np}) \qquad (5)$$

$$r_{ij} = \rho\sqrt{1 - cos^2(\theta - \varphi_{ij} + \pi)\, cos^2(\alpha_{ij})} \qquad (6)$$

$$r_{np} = \rho\sqrt{1 - cos^2(\theta - \varphi_{np} + \pi)\, cos^2(\alpha_{np})} \qquad (7)$$

The intensities of the (i,j) and (n,p) laser beams at the image plane can thus be expressed, respectively, as:

$$I_{i,j}(\rho,\theta) = I_0 \frac{{w_0}^2}{w^2(z_{ij}+\Delta z_{ij})}\ exp\left[-2\frac{\rho^2\left(1-cos^2(\theta-\varphi_{ij}+\pi)\ cos^2(\alpha_{ij})\right)}{w^2(z_{ij}+\Delta z_{ij})}\right] \qquad (8)$$

$$I_{n,p}(\rho,\theta) = I_0 \frac{{w_0}^2}{w^2(z_{np}+\Delta z_{np})}\ exp\left[-2\frac{\rho^2\left(1-cos^2(\theta-\varphi_{np}+\pi)\ cos^2(\alpha_{np})\right)}{w^2(z_{np}+\Delta z_{np})}\right] \qquad (9)$$

Since the combined laser beams are mutually incoherent, the total intensity $I_T(\rho,\theta)$ measured at the image plane is given by the sum of the intensities of all laser beams emitted from the two source planes.

$$I_T(\rho,\theta) = \textstyle\sum_{i,j} I_{i,j}(\rho,\theta) + \sum_{n,p} I_{n,p}(\rho,\theta) \qquad (10)$$

$z_{ij} = \sqrt{{x_i}^2 + {y_j}^2 + {h_{10}}^2}$ and $z_{np} = \sqrt{{x_n}^2 + {y_p}^2 + {h_{20}}^2}$ are the distances from the position of the (i,j) laser beam in source plane 1 and the (n,p) laser beam in source plane 2, respectively, to the origin of the image plane. Likewise, $\varphi_{ij}$ and $\varphi_{np}$ denote the polar angles locating the (i,j) laser beam in source plane 1 and the (n,p) laser beam in source plane 2, respectively.

Having determined the intensity at the image plane defined by $z_1 = h_{10}$, we now focus on calculating the total intensity $I_T(X,Y,h)$ at an arbitrary image plane defined by $z_1 = h_1$, where $h_1$ is the distance between source plane 1 and the image plane $z_1 = h_1$, as illustrated in Figure 3. For source plane 2, the corresponding image plane is defined by $z_2 = h_2$, such that $h_{10} + h_{20} = h_1 + h_2$ which represents the constant distance between source planes 1 and 2.

Let $O'$ and $O''$ be the intersection points of the plane defined by $z_1 = h_1$ with the lines $z_{ij}$ and $z_{np}$ respectively. The coordinates of $O'$ in the coordinate system of source plane 1 are $(x_{0i}, y_{0j}, h_1)$, while the coordinates of $O''$ in the coordinate system of source plane 2 are $(x_{0n}, y_{0p}, h_2)$, as illustrated in

Figure 3. For $O'$, $x_{0i} = x_i\left(1-\frac{h_1}{h_{10}}\right)$ and $y_{0j} = y_j\left(1-\frac{h_1}{h_{10}}\right)$. And for $O''$ , $x_{0n} = x_n\left(1-\frac{h_2}{h_{20}}\right)$ and $y_{0p} = y_p\left(1-\frac{h_2}{h_{20}}\right)$.

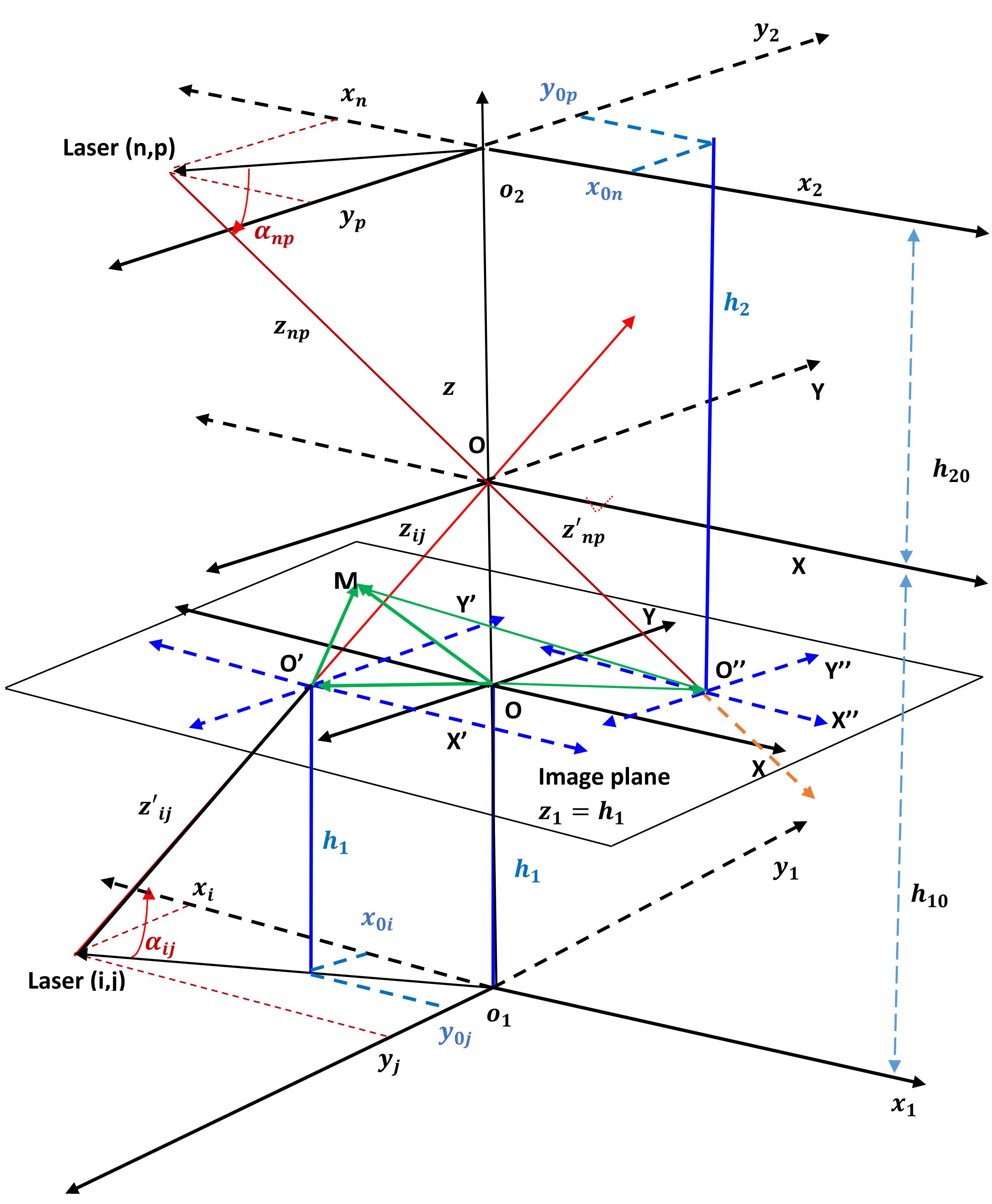

**Fig 3 : Geometrical determination of $\Delta z_{ij}{}'$ , $r_{ij}{}'$ , $\Delta z_{np}{}'$ , $r_{np}{}'$ at the plane $z_1 = h_1$**

By choosing $(O', X', Y')$ as the reference coordinate system for calculating the optical intensity distribution of the (i,j) laser beam, equations similar to (4) and (6) are obtained. Thus, we obtain

$$\Delta z'_{i,j} = \rho' \cos(\theta' - \varphi_{ij} + \pi) \cos(\alpha_{ij}) \quad \textbf{(11)}$$

$$r'_{i,j} = \rho' \sqrt{1 - cos^2(\theta' - \varphi_{ij} + \pi)\, cos^2(\alpha_{ij})} \quad \textbf{(12)}$$

Similarly, for the (n,p) laser beam, by choosing $(O'', X'', Y'')$ as the reference coordinate system, we obtain

$$\Delta z'_{n,p} = \rho'' \cos(\theta'' - \varphi_{np} + \pi) \cos(\alpha_{np}) \quad \textbf{(13)}$$

$$r'_{n,p} = \rho'' \sqrt{1 - cos^2(\theta'' - \varphi_{np} + \pi)\, cos^2(\alpha_{np})} \quad \textbf{(14)}$$

To calculate the total intensity of all laser beams originating from source plane 1 at the plane $z_1 = h_1$, and with respect to the reference frame $(O, X, Y)$, the coordinates $(\rho', \theta')$ must be expressed as functions of $X$ and $Y$. Since $\overrightarrow{OM} = \overrightarrow{OO'} + \overrightarrow{O'M}$ as illustrated in Figure 3, it follows that $X' = X - x_{0i}$ and $Y' = Y - y_{0j}$ and therefore $\rho' = \sqrt{(X - x_{0i})^2 + (Y - y_{0j})^2}$ and $\theta' = arctang\left(\frac{Y - y_{0j}}{X - x_{0i}}\right)$. The intensity of the (i,j) laser beam at the plane $z_1 = h_1$ is then given by

$$I_{i,j}(X, Y, h_1) = I_0 \frac{{w_0}^2}{w^2(z'_{i,j} + \Delta z'_{i,j})}\ exp\left[-2\frac{\rho'^2\left(1 - cos^2(\theta' - \varphi_{ij} + \pi)\, cos^2(\alpha_{ij})\right)}{w^2(z'_{i,j} + \Delta z'_{i,j})}\right] \quad \textbf{(15)}$$

Where $z'_{i,j} = \frac{h_1}{h_{10}} z_{ij}$ represents the distance measured along the propagation line connecting the (i,j) laser source in source plane 1 to the point $O'$.

Similarly, to calculate the total intensity of all laser beams originating from source plane 2 at the plane $z_1 = h_1$ or equivalently $z_2 = h_2$, with respect to the reference frame $(O, X, Y)$, the coordinates $(\rho'', \theta'')$ must be expressed as functions of $X$ and $Y$. Since $\overrightarrow{OM} = \overrightarrow{OO''} + \overrightarrow{O''M}$ as illustrated in Figure 3, it follows that $X'' = X - x_{0n}$ and $Y'' = Y - y_{0p}$ and therefore $\rho'' = \sqrt{(X - x_{0n})^2 + (Y - y_{0p})^2}$ and $\theta'' = arctang\left(\frac{Y - y_{0p}}{X - x_{0n}}\right)$. The intensity of the (n,p) laser beam at the plane $z_2 = h_2$ is then given by

$$I_{n,p}(X, Y, h_1) = I_0 \frac{{w_0}^2}{w^2(z'_{n,p} + \Delta z'_{n,p})}\ exp\left[-2\frac{\rho''^2\left(1 - cos^2(\theta'' - \varphi_{np} + \pi)\, cos^2(\alpha_{np})\right)}{w^2(z'_{n,p} + \Delta z'_{n,p})}\right] \quad \textbf{(16)}$$

Where $z'_{n,p} = \frac{h_2}{h_{20}} z_{np}$ represents the distance measured along the propagation line connecting the (n,p) laser source in source plane 2 to the point $O''$.

Because the combined laser beams are mutually incoherent, the total intensity measured in the plane $z_1 = h_1$ is simply the sum of the individual intensities contributed by all beams from source planes 1 and 2

$$I_T(X, Y, h_1) = \sum_{i,j} I_{i,j}(X, Y, h_1) + \sum_{n,p} I_{n,p}(X, Y, h_1) \quad \textbf{(17)}$$

The previous calculations were carried out for unfocused laser beams. To achieve very high intensities at the focal plane, it is necessary to focus the laser beams originating from source planes 1 and 2 onto the focal point O. Furthermore, to maximize the beam-combining efficiency, each laser beam from source plane 1 is focused by a lens with focal length $f_{ij} = z_{ij}$, while each laser beam from source plane 2 is focused by a lens with focal length $f_{np} = z_{np}$.

Since focused Gaussian beams retain the same Gaussian profile, it is sufficient to determine the new beam waist $w(z)$ and the new radius of curvature $R(z)$ of the focused beam in order to calculate the optical field and intensity at any image plane.

After focusing a Gaussian beam with a minimum waist $w_0$ by a lens of focal length $f$, the radius of curvature $R(z)$ and the beam waist $w(z)$ of the focused beam are given by **[1]**:

$$R(z) = \frac{\left(1-\frac{z}{f}\right)^2+\left(\frac{z}{z_0}\right)^2}{\frac{-1}{f}+z\left(\frac{1}{f^2}+\frac{1}{z_0^2}\right)} \quad \textbf{(18)}$$

$$w(z) = \sqrt{\frac{\lambda}{\pi}}\sqrt{\frac{\left(1-\frac{z}{f}\right)^2+\left(\frac{z}{z_0}\right)^2}{\frac{1}{z_0}}} \quad \textbf{(19)}$$

Because each laser beam originating from source planes 1 and 2 is focused by a lens of focal length $f_{ij} = z_{ij}$ and $f_{np} = z_{np}$, respectively, the beam waist and radius of curvature of the focused (i,j) laser beam at the propagation distance $z_1 = z'_{i,j} + \Delta z'_{i,j}$ can be expressed as:

$$R\left(z'_{i,j} + \Delta z'_{i,j}\right) = \frac{\left(1-\frac{z'_{i,j}+\Delta z'_{i,j}}{z_{ij}}\right)^2+\left(\frac{z'_{i,j}+\Delta z'_{i,j}}{z_0}\right)^2}{\frac{-1}{z_{ij}}+\left(z'_{i,j}+\Delta z'_{i,j}\right)\left(\frac{1}{z_{ij}{}^2}+\frac{1}{z_0^2}\right)} \quad \textbf{(20)}$$

$$w\left(z'_{i,j} + \Delta z'_{i,j}\right) = \sqrt{\frac{\lambda}{\pi}}\sqrt{\frac{\left(1-\frac{z'_{i,j}+\Delta z'_{i,j}}{z_{ij}}\right)^2+\left(\frac{z'_{i,j}+\Delta z'_{i,j}}{z_0}\right)^2}{\frac{1}{z_0}}} \quad \textbf{(21)}$$

Similarly, the beam waist and radius of curvature of the focused (n,p) laser beam at a distance $z_2 = z'_{n,p} + \Delta z'_{n,p}$ are given by:

$$R\left(z'_{n,p} + \Delta z'_{n,p}\right) = \frac{\left(1-\frac{z'_{n,p}+\Delta z'_{n,p}}{z_{n,p}}\right)^2+\left(\frac{z'_{n,p}+\Delta z'_{n,p}}{z_0}\right)^2}{\frac{-1}{z_{n,p}}+\left(z'_{n,p}+\Delta z'_{n,p}\right)\left(\frac{1}{z_{np}{}^2}+\frac{1}{z_0^2}\right)} \quad \textbf{(22)}$$

$$w\left(z'_{n,p} + \Delta z'_{n,p}\right) = \sqrt{\frac{\lambda}{\pi}}\sqrt{\frac{\left(1-\frac{z'_{n,p}+\Delta z'_{n,p}}{z_{n,p}}\right)^2+\left(\frac{z'_{n,p}+\Delta z'_{n,p}}{z_0}\right)^2}{\frac{1}{z_0}}} \quad \textbf{(23)}$$

Hence, the total intensity distribution in the image plane $z_1 = h_1$, given by Eq. (17), can be obtained by substituting the focused-beam expressions of $w\left(z'_{i,j} + \Delta z'_{i,j}\right)$ and $w\left(z'_{n,p} + \Delta z'_{n,p}\right)$ into Eqs. (15) and (16), respectively.

The parameters used in these simulations are listed in Table 1.

| Laser wavelengths | $\lambda = 0.8\ \mu m$ |
|---|---|
| Minimum beam waist | $w_0 = 0.01\ m$ |
| Distance between adjacent laser beams | 0.04 m |
| Source plane 1–focal point distance | $h_{10} = 1m$ |
| Source plane 2–focal point distance | $h_{20} = 1m$ |
| Number of laser beams in each source plane | A square array of 121 laser beams at the source plane. |

**Table 1: Parameters Used in the Simulations**

Figure 4 shows the spatial evolution of the combined intensity along the focusing axis (Z). The figures on the left were obtained using two source planes (4π focusing), whereas the figures on the right were obtained using a single source plane (2π focusing). We observe that the combined intensity exhibits a Gaussian-like profile that extends longitudinally over a distance of approximately 360λ, with a full width at half maximum (FWHM) of several wavelengths, as illustrated in Figure 5.

The intensity obtained with two source planes is twice that obtained with a single source plane. However, the beam waist w(z) and the FWHM remain unchanged for both 2π and 4π focusing configurations, as shown in Figure 5.

Our simulations also indicate that the longitudinal focal length strongly depends on the numerical aperture (NA) of the system. Indeed, it is possible to achieve longitudinal focal lengths extending over several centimeters with low NA values. For example, by focusing the lasers onto a focal plane located 10 m from the source planes ($h_{10}$ = $h_{20}$ = 10 m), a longitudinal focal region extending over approximately 30,000λ can be obtained.

We further observe that at Z = ∓300λ, the wavefront of the combined intensity exhibits a square-shaped profile with nearly uniform intensity. This feature could be advantageous for inertial confinement fusion (ICF) applications. In this context, the inverse problem, which consists of determining the source-laser parameters required to produce a desired combined-intensity wavefront, represents an interesting research topic. For instance, it would be valuable to determine the source-plane laser parameters necessary to generate a spherical wavefront suitable for spherical targets used in inertial confinement fusion.

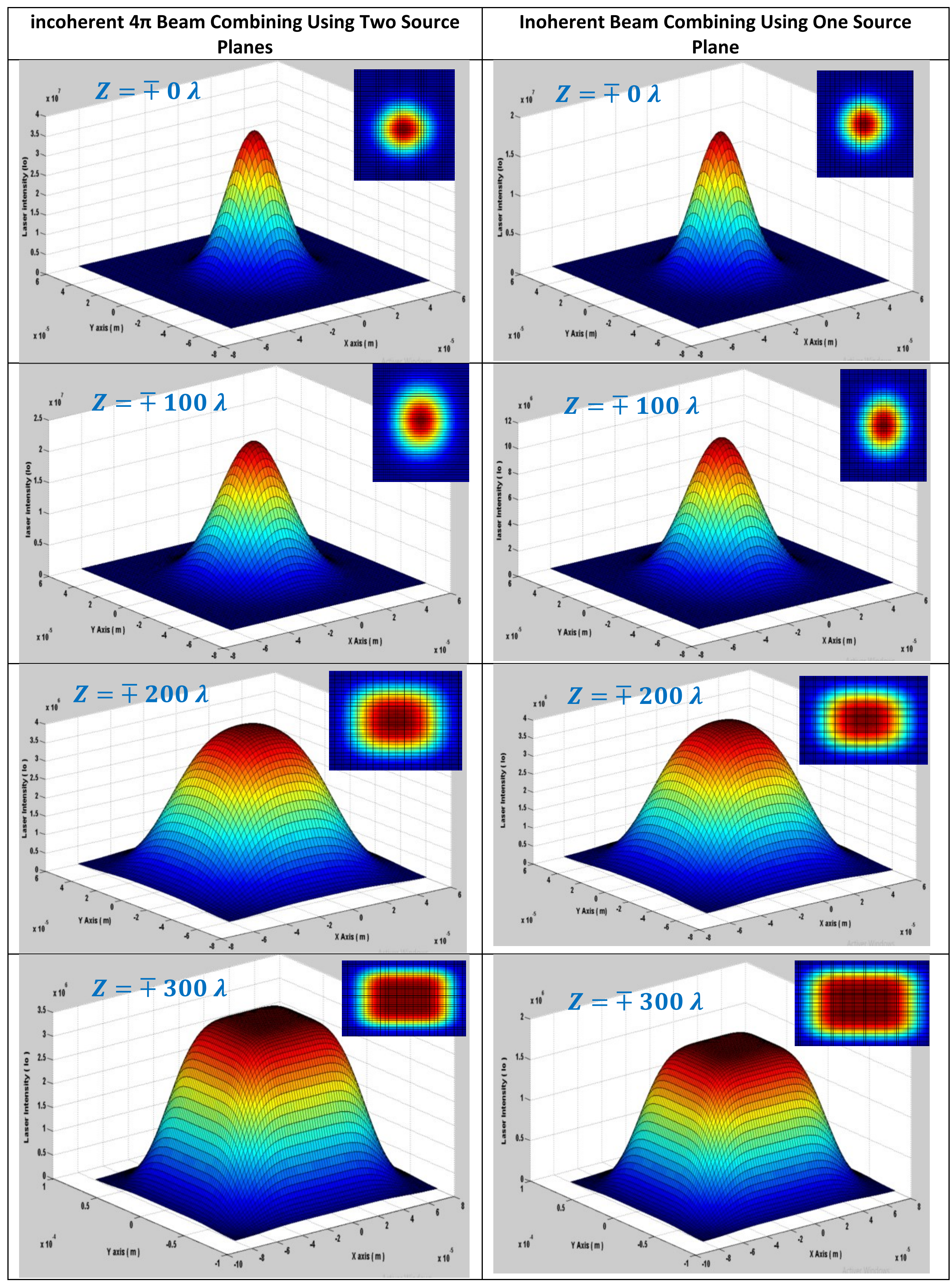

**Fig. 4. Spatial evolution of the combined intensity obtained through incoherent beam combining using two source planes (left panels) and a single source plane (right panels).**

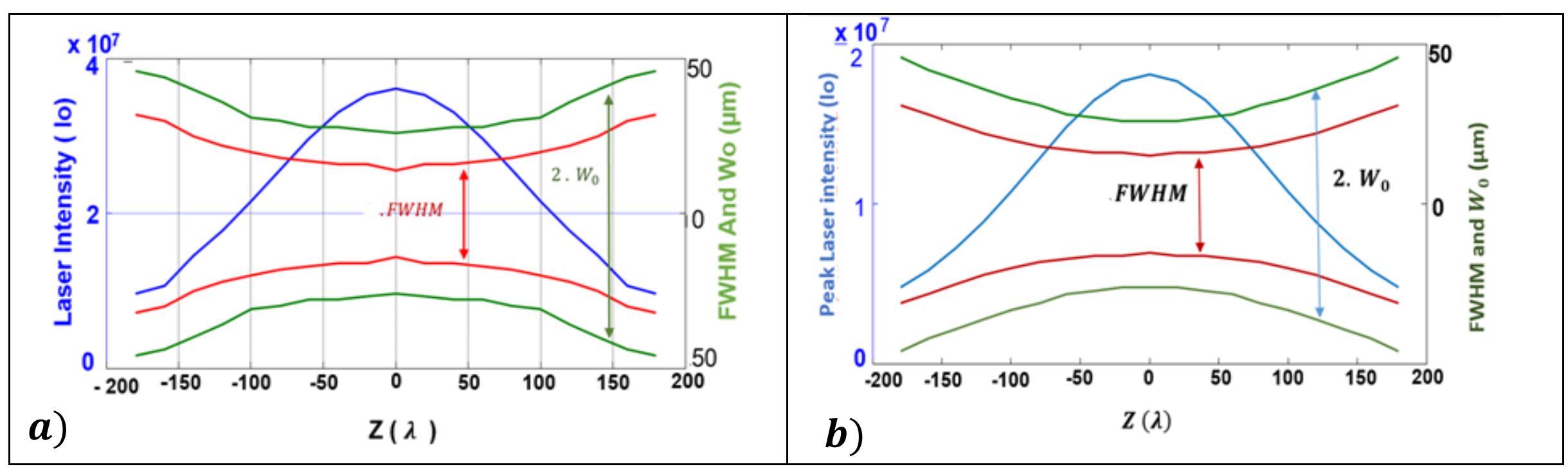

**Fig. 5. Evolution of the peak combined intensity, W(z), and FWHM along the focusing axis for (a) two source planes and (b) one source plane under incoherent beam combining.**

## 2- Coherent Beam Combining

In the case of incoherent beam combining, the phase, spectrum, and polarization states of the combined lasers can be arbitrary. In contrast, coherent beam combining requires all laser beams to be phase-locked so that they are in phase at the focal point.

In this study, we have chosen a linear polarization initially aligned along the x-axis. However, in our configuration, each laser beam in the source plane must be directed toward the image focal point. Since the electric field is always perpendicular to the propagation direction, the electric field cannot remain exclusively oriented along the x-direction. Instead, it generally possesses components along the three spatial directions (x, y, and z).

Nevertheless, it is possible to reduce the polarization to only two components by rotating each laser around its propagation axis. In this case, the electric-field amplitude of the laser beam (i,j) in source plane 1 can be expressed as follows **[1]**:

$$\overrightarrow{E_{0ij}} = E_0\left(\frac{h_{10}}{\sqrt{x_i^2+h_{10}^2}}\vec{i} + \frac{x_i}{\sqrt{x_i^2+h_{10}^2}}\overrightarrow{e_z}\right) \qquad \textbf{(24)}$$

Thus, the electric field of the laser beam (i,j) at the image plane $z_1 = h_{10}$ can be written as:

$$\overrightarrow{E_{ij}}(\rho,\theta,h_{10}) = \sqrt{I_0}\left(\frac{h_{10}}{\sqrt{x_i^2+h_{10}^2}}\vec{i} + \frac{x_i}{\sqrt{x_i^2+h_{10}^2}}\overrightarrow{e_z}\right)\frac{w_0}{w(z_{ij}+\Delta z_{ij})}\, exp\left[-\frac{{r_{ij}}^2}{w^2(z_{ij}+\Delta z_{ij})}\right]\, exp\left[i\left(k(z_{ij}+\Delta z_{ij}) - arctan\left(\frac{z_{ij}+\Delta z_{ij}-f_{ij}}{z_{0ij}}\right) + \frac{k{r_{ij}}^2}{2R(z_{ij}+\Delta z_{ij})}\right)\right] \qquad \textbf{(25)}$$

Here, $z_{0ij} = \frac{\pi}{\lambda}w^2(f_{ij})$ is the new Rayleigh range of the (i,j) laser beam after focusing, where $w(f_{ij})$ denotes the minimum beam waist of the focused laser beam (i,j). Since $z_{ij} = f_{ij}$, the Gouy phase becomes $-arctan\left(\frac{\lambda\,\Delta z_{ij}}{\pi\, w^2(f_{ij})}\right)$, Since $\lambda = 0.8\ \mu m$ and the value of $\Delta z_{ij}$ necessary for constructive interference is only a few micrometers, and that $w(f_{ij})$ is approximately 25 µm for the parameters used in this work, the Gouy phase term can be neglected. Similarly, the term $\frac{k r_{ij}^2}{2R(z_{ij}+\Delta z_{ij})}$ can be neglected because, at the focal plane, $R(z_{ij}+\Delta z_{ij})$ tends to infinity. Furthermore, because all laser beams are phase-matched at the focal point, the term $k\, z_{ij}$ is the same for every laser beam (i,j). With these approximations, and by replacing $\cos(\alpha_{ij})$ with $\sin(\phi_{ij})$, where $\phi_{ij}$ denotes the viewing angle of laser beam (i,j) from the image focal point (see Fig. 2), the electric field of laser beam (i,j) at the image plane becomes:

$$\overrightarrow{E_{ij}}(\rho,\theta,h_{10}) = \sqrt{I_0}\left(\frac{h_{10}}{\sqrt{x_i^2+h_{10}^2}}\vec{\imath} + \frac{x_i}{\sqrt{x_i^2+h_{10}^2}}\overrightarrow{e_z}\right)\frac{w_0}{w(z_{ij}+\Delta z_{ij})}\, exp\left[-\frac{r_{ij}^2}{w^2(z_{ij}+\Delta z_{ij})}\right] exp\left(i2\pi\, \rho\, \cos(\theta - \varphi_{ij}+\pi)\,\frac{\sin(\phi_{ij})}{\lambda}\right) \quad \textbf{(26)}$$

For the electric field of laser beam (n,p) originating from source plane 2, the electric-field component along the (z)-axis is negative, such that $\overrightarrow{e_{2z}} = -\overrightarrow{e_{1z}} = \overrightarrow{-e_z}$, as illustrated in Fig. 2. This sign difference arises because the electric field is evaluated at the image plane. Following the same reasoning as that used for laser beam (i,j), the electric field of laser beam (n,p) at the image plane, $z_1 = h_{10}$ or equivalently $z_2 = h_{20}$ , can be written as:

$$\overrightarrow{E_{np}}(\rho,\theta,h_{10}) = \sqrt{I_0}\left(\frac{h_{20}}{\sqrt{x_n^2+h_{20}^2}}\vec{\imath} - \frac{x_n}{\sqrt{x_n^2+h_{20}^2}}\overrightarrow{e_z}\right)\frac{w_0}{w(z_{np}+\Delta z_{np})}\, exp\left[-\frac{r_{np}^2}{w^2(z_{np}+\Delta z_{np})}\right] exp\left(i2\pi\, \rho\, \cos(\theta - \varphi_{np}+\pi)\,\frac{\sin(\phi_{np})}{\lambda}\right) \quad \textbf{(27)}$$

The total electric field produced by all the (i,j) laser beams and the (n,p) laser beams at the focal plane is given by

$$\overrightarrow{E_T} = \sum_{i,j}\overrightarrow{E_{ij}}(\rho,\theta,h_{10}) + \sum_{n,p}\overrightarrow{E_{np}}(\rho,\theta,h_{10}) \quad \textbf{(28)}$$

The total intensity at the image plane is then expressed as

$$I_T(\rho,\theta,h_{10}) = \left|\overrightarrow{E_T}(\rho,\theta,h_{10})\,\overrightarrow{E_T}^{*}(\rho,\theta,h_{10})\right| \quad \textbf{(29)}$$

It follows from Eqs. (26), (27), and (28) that, when the two source planes are identical and $h_{10} = h_{20}$, the total electric field at the focal plane is polarized exclusively along the (X)-direction. Conversely, when the laser beams originating from the two source planes exhibit a phase difference of $\pi$, the total electric field is polarized along the (Z)-direction.

After calculating the total electric field and intensity at the image plane defined by $z_1 = h_{10}$, we now determine the electric field and the total intensity at an arbitrary plane defined by $z_1 = h_1$, as illustrated in Figure 3.

Since all the (i,j) laser beams from source plane 1 are in phase at the focal point (O) on the image plane $z_1 = h_{10}$, their phase at point (O'), defined as the intersection of the axis of the (i,j) laser beam with the image plane $z_1 = h_1$, will be different. The corresponding phase shift is given by $-\frac{2\pi}{\lambda}(z_{ij} - z_{ij}')$. However, the polarization of the (i,j) laser beam remains unchanged because the two planes $z_1 = h_{10}$ and $z_1 = h_1$ are parallel. Therefore, the electric field at the image plane defined by $z_1 = h_1$ can be written as

$$\overrightarrow{E_{ij}}(\rho,\theta,h_1) = \sqrt{I_0}\left(\frac{h_{10}}{\sqrt{x_i^2+h_{10}^2}}\vec{\imath} + \frac{x_i}{\sqrt{x_i^2+h_{10}^2}}\overrightarrow{e_z}\right)\frac{w_0}{w(z_{ij}'+\Delta z_{ij}')}\, exp\left[-\frac{r_{ij}^2}{w^2(z_{ij}'+\Delta z_{ij}')}\right] exp\left[i\left(k\big((z_{ij}' - z_{ij}) + \Delta z_{ij}'\big) - arctan\left(\frac{z_{ij}'+\Delta z_{ij}'-f_{ij}}{z_{0ij}}\right) + \frac{kr_{ij}^2}{2R(z_{ij}'+\Delta z_{ij}')}\right)\right] \qquad \textbf{(30)}$$

Following the same procedure for the (n,p) laser beam originating from source plane 2, the corresponding electric field is obtained as

$$\overrightarrow{E_{np}}(\rho,\theta,h_1) = \sqrt{I_0}\left(\frac{h_{20}}{\sqrt{x_n^2+h_{20}^2}}\vec{\imath} - \frac{x_n}{\sqrt{x_n^2+h_{20}^2}}\overrightarrow{e_z}\right)\frac{w_0}{w(z_{np}'+\Delta z_{np}')}\exp\left[-\frac{r_{ij}^2}{w^2(z_{np}'+\Delta z_{np}')}\right]\exp\left[i\left(k\left((z_{np}' - z_{np}) + \Delta z_{np}'\right) - \arctan\left(\frac{z_{np}'+\Delta z_{np}'-f_{np}}{z_{0np}}\right) + \frac{kr_{np}^2}{2R(z_{np}'+\Delta z_{np}')}\right)\right] \qquad \textbf{(31)}$$

The total electric field at the image plane $z_1 = h_1$, resulting from the superposition of all (i,j) and (n,p) laser beams, can be expressed as

$$\overrightarrow{E_T} = \textstyle\sum_{i,j}\overrightarrow{E_{ij}}(\rho,\theta,h_1) + \sum_{n,p}\overrightarrow{E_{np}}(\rho,\theta,h_1) \qquad \textbf{(32)}$$

The resulting total intensity at the image plane can therefore be expressed as

$$I_T(\rho,\theta,h_1) = \left|\overrightarrow{E_T}(\rho,\theta,h_1)\,\overrightarrow{E_T}^{*}(\rho,\theta,h_1)\right| \qquad \textbf{(33)}$$

Figure 6 illustrates the spatial evolution of the combined intensity along the focusing axis (Z). The panels on the right correspond to the coherent combination of single source plane (2π focusing), whereas the panels on the left correspond to the coherent combination of two source planes (4π focusing).

In the vicinity of the focal plane, a central Gaussian-shaped peak can be observed, surrounded by secondary lobes of lower intensity aligned along the (x)- and (y)-axes. These lobes exhibit a sinc-like profile, corresponding to the Fourier transform of a two-dimensional rectangular function. This behavior arises because the combined laser beams are arranged in a square configuration at the source planes.

The peak intensity obtained with two source planes is four times greater than that obtained with a single source plane. However, the waist and the full width at half maximum (FWHM) of the combined intensity remain unchanged for both 4π and 2π focusing configurations (see Figure 7), despite the fact that the numerical aperture in the 4π focusing configuration is twice that of the 2π configuration.

It can also be observed that the depth of focus is shorter in the 4π focusing configuration than in the 2π configuration, as shown in Figure 6. Indeed, the combined intensity profile in the 4π configuration is no longer Gaussian at $z = \pm 10\ \lambda$, whereas in the 2π configuration it retains its Gaussian shape up to approximately $z = \pm 20\ \lambda$.

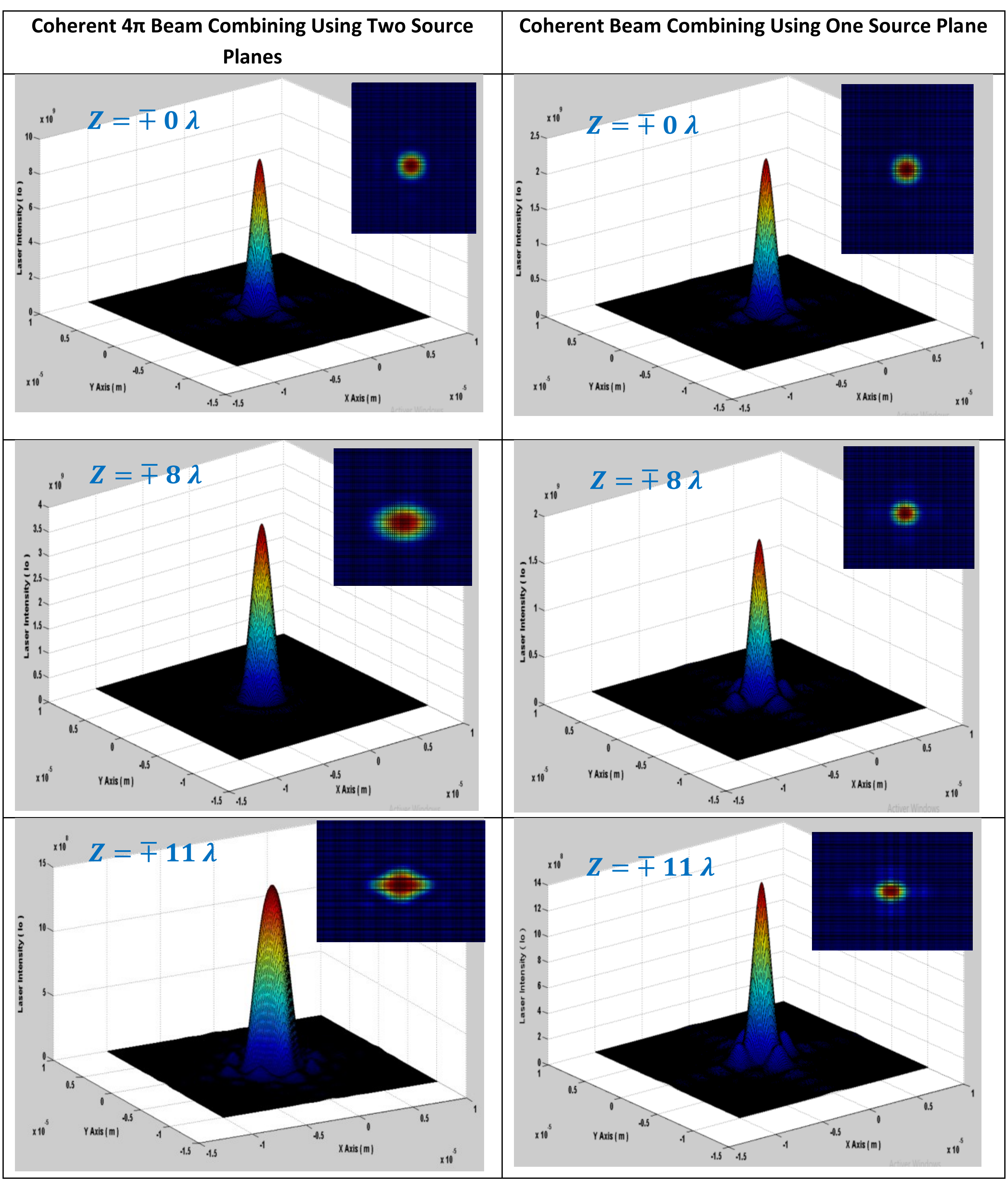

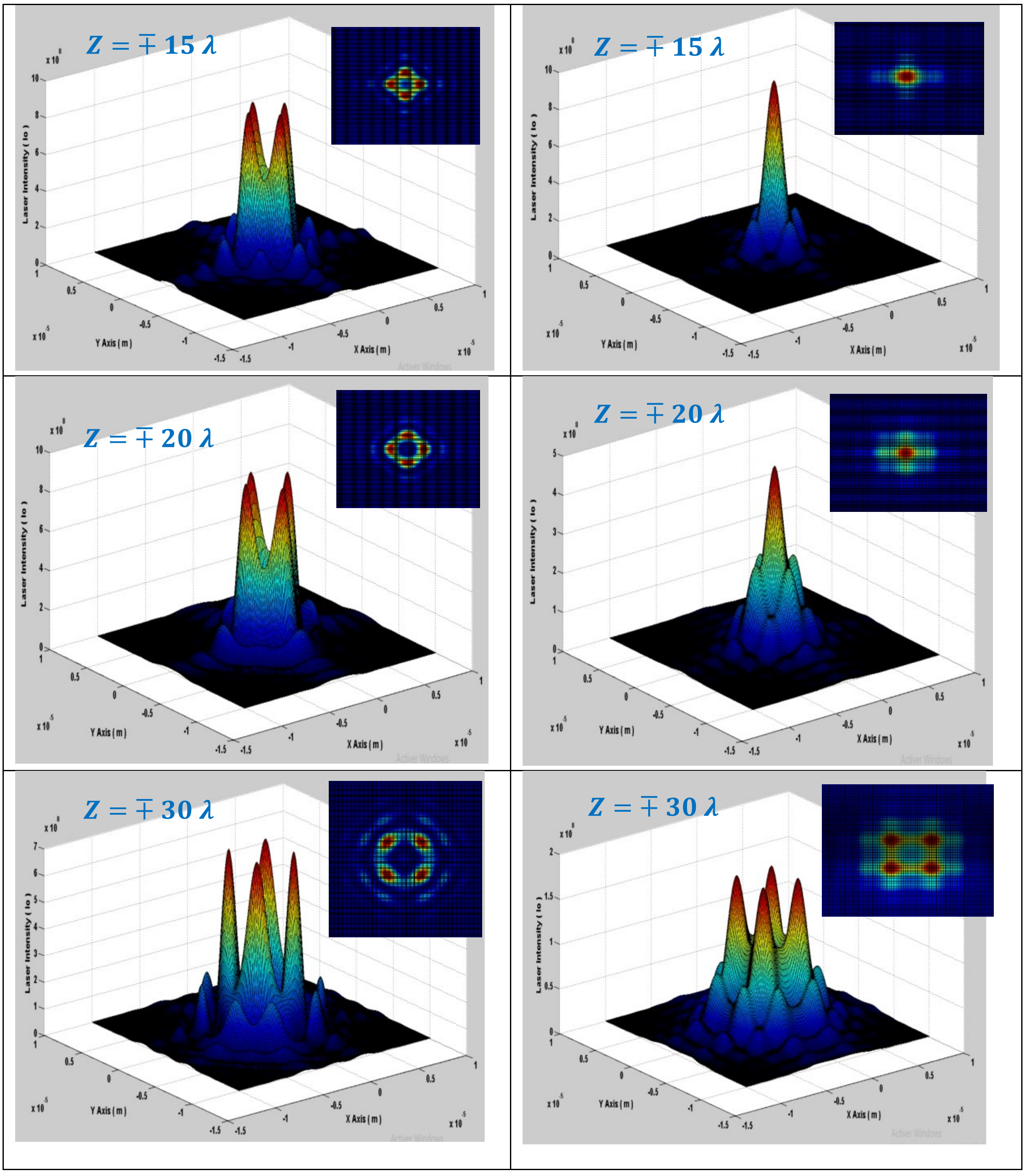

**Fig 6. Spatial evolution of the combined intensity obtained through coherent 4π beam combining using two source planes (left panels) and coherent 2π beam combining using one source plane (right panels).**

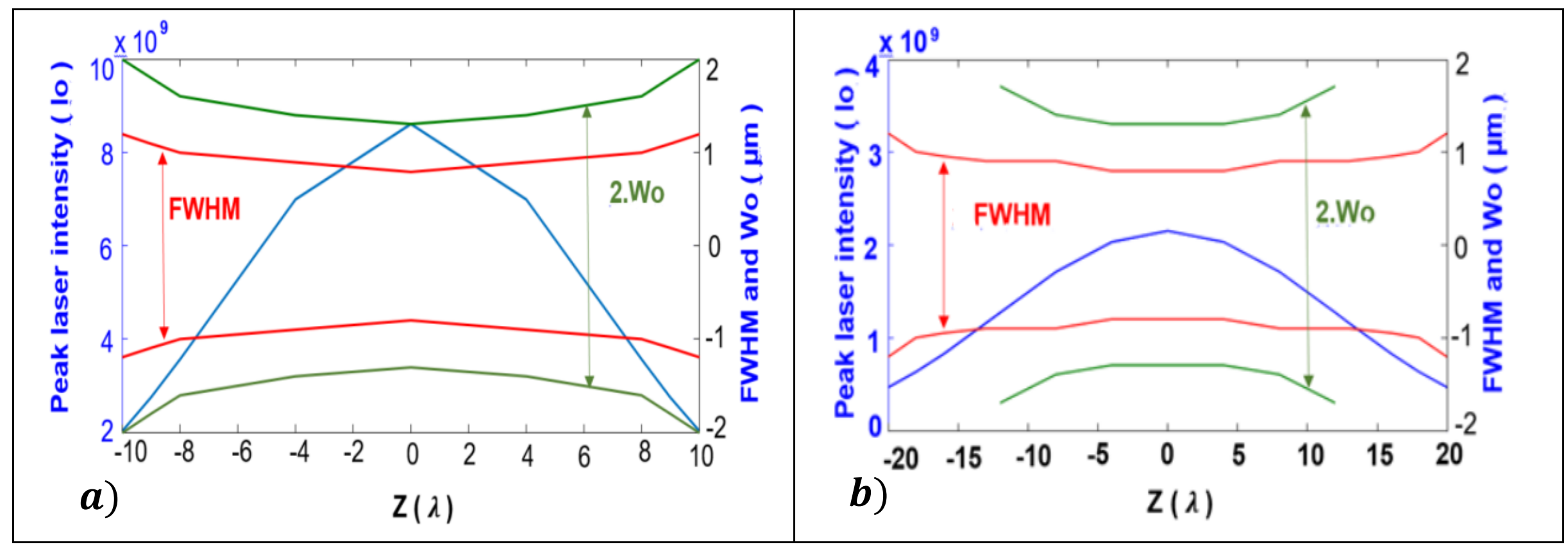

**Fig 7. Variation of the peak combined intensity, beam waist W(z), and full width at half maximum (FWHM) along the focusing axis. (a) Coherent 4π beam combining with two source planes. (b) Coherent 2π beam combining with one source plane.**

It can be observed that, in the case of coherent beam combining using a single source plane, the combined intensity profile is no longer the Fourier transform of a square aperture when the numerical aperture becomes large $NA > 0.82$. As shown in Figure 8(a), obtained for $h_{10} = h_{20} = 0.14\,m$, the central lobe begins to exhibit noticeable distortion. For $h_{10} = h_{20} = 0.05\,m$, corresponding to a numerical aperture of $NA = 0.97$, the resulting intensity distribution exhibits three central lobes, as illustrated in Figure 8(b).

In contrast, for coherent beam combining using two source planes, the central lobe of the combined intensity retains an approximately Gaussian profile with an ellipsoidal base, even at large numerical apertures $NA > 0.82$. However, the overall intensity distribution is no longer the Fourier transform of a square aperture. Instead, it evolves toward a pattern resembling the radiation intensity distribution of a dipole emitter, as shown in Figure 9.

For coherent 4π focusing with linearly polarized laser beams and a focal distance of $h_{10} = h_{20} = 0.1\,m$, corresponding to a numerical aperture of $NA = 0.895$, the focal spot area is $FWHMx * FWHMy = 0.5\,\lambda * 0.425\,\lambda = 0.2125\,\lambda^2$, The corresponding focal volume is $FWHMx * FWHMy * FWHMz = 0.5\,\lambda * 0.425\,\lambda * 0.4\,\lambda = 0.085\,\lambda^3$.

The focal spot area obtained with 4π focusing, $0.2125\,\lambda^2$, is smaller than that reported for linearly polarized focusing $0.26\,\lambda^2$, but remains larger than that achieved with radially polarized focusing $0.16\,\lambda^2$ **[10]**. Therefore, it would be of considerable interest to investigate coherent beam combining using radially polarized laser beams, as such a configuration may further improve the focusing performance and reduce the focal spot size.

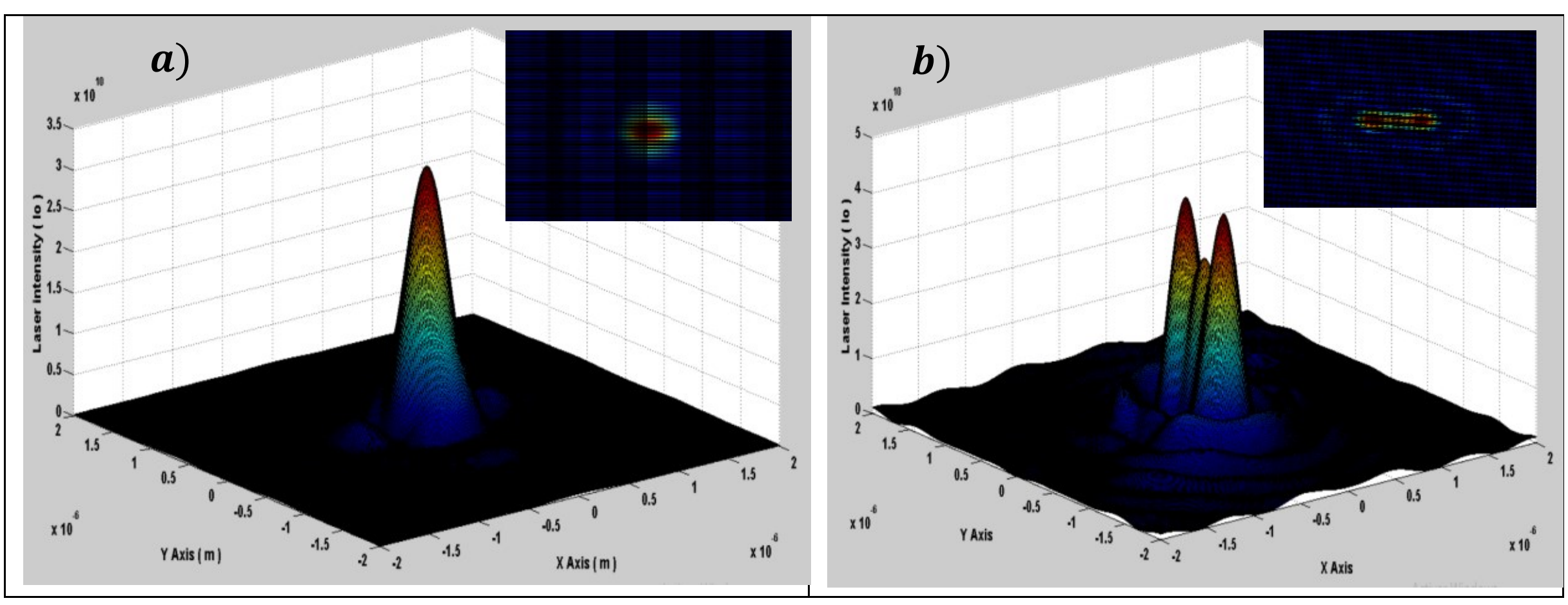

**Fig 8. Coherent 2π beam combining using a single source plane at high numerical aperture: (a) $\boldsymbol{h_{10} = h_{20} = 0.14\ m}$; (b) $\boldsymbol{h_{10} = h_{20} = 0.05\ m}$.**

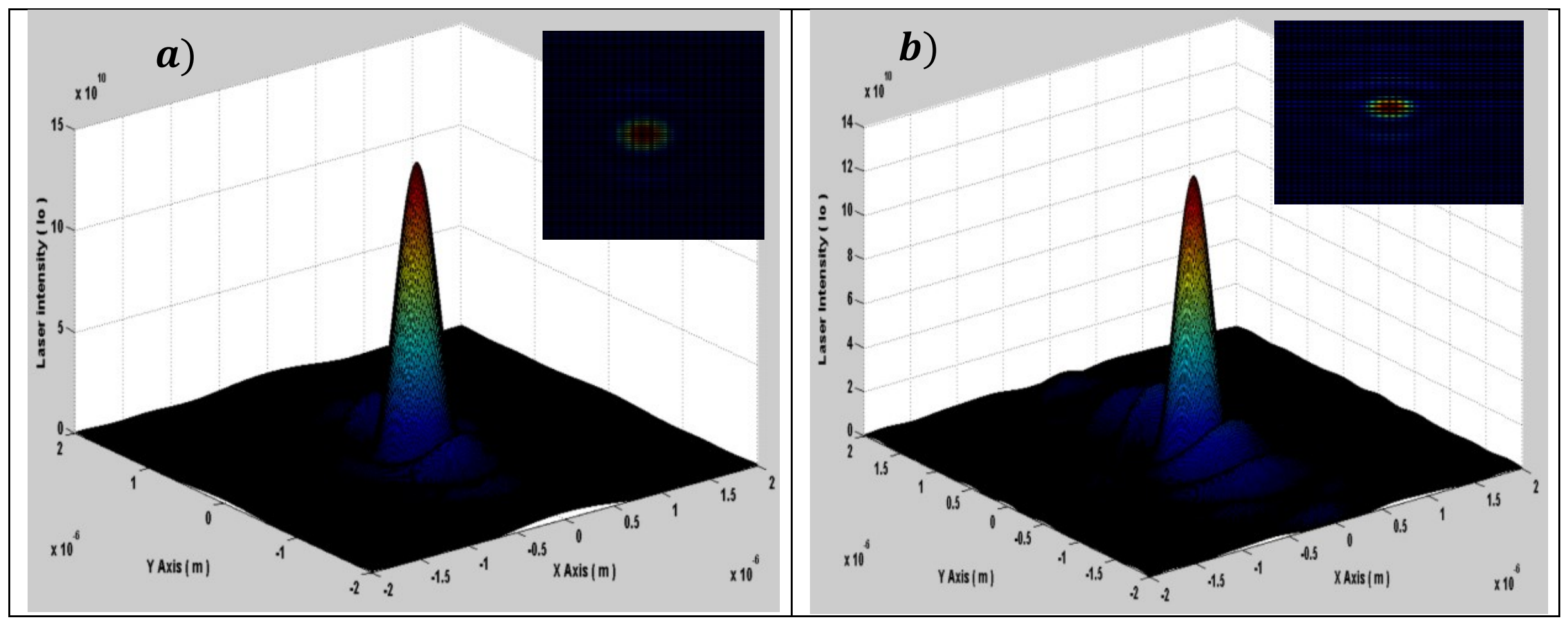

**Fig 9. Intensity distributions obtained by coherent 4π beam combining with two source planes at high numerical aperture: (a) $\boldsymbol{h_{10} = h_{20} = 0.1\ m}$; (b) $\boldsymbol{h_{10} = h_{20} = 0.05\ m}$.**

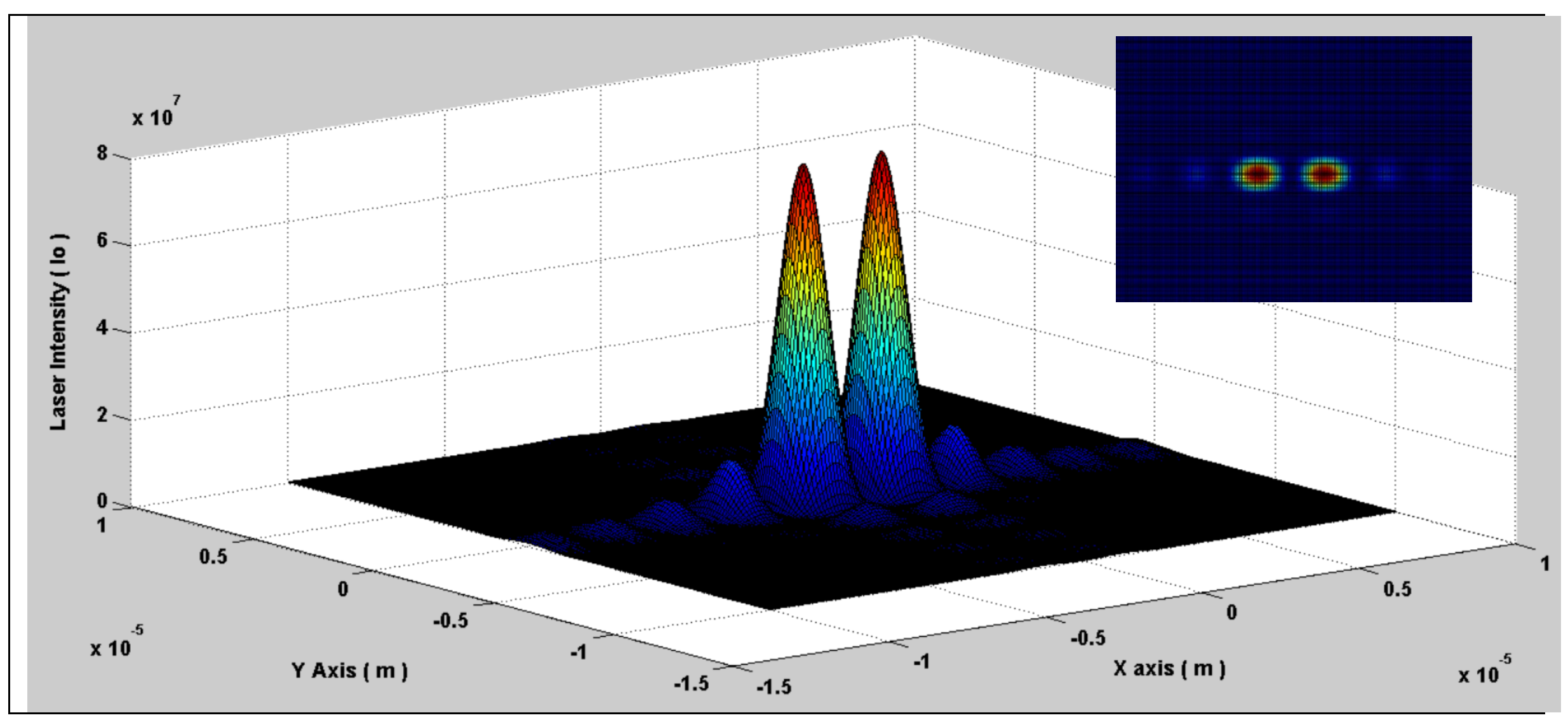

**Figure 10. Intensity distribution obtained by coherent 4π beam combining with two source planes and a phase difference of $\pi$ between the laser beams from the two source planes.**

Figure 10 was obtained with a phase difference of $\pi$ between the laser beams originating from the two source planes. As explained previously, this phase relationship results in a total electric field at the focal plane that is polarized exclusively along the (z)-direction.

The figure corresponds to the case $h_{10} = h_{20} = 1\ m$. It can be observed that the intensity vanishes at the center of the focal plane because the total resulting electric field at point (O) is zero. Instead of a single central maximum, two central lobes are formed, followed by secondary lobes extending along the (x)- and (y)-directions.

## Conclusion

In this work, we investigated the coherent and incoherent combining of Gaussian laser beams in a 4π configuration involving two opposing source planes. This study enabled us to determine the total combined electric field and intensity at every point in space.

The simulation results show that, in the case of coherent beam combining, the focal spot area remains the same for both the 4π and 2π configurations. However, the intensity at the focal point is four times higher in the 4π configuration than in the 2π configuration. In addition, the depth of focus obtained with 4π beam combining is shorter than that achieved with 2π beam combining. For large numerical apertures $NA > 0.82$, the combined intensity distribution is no longer described by the Fourier transform of a square aperture. In the 2π configuration, multiple lobes appear in the focal plane, whereas in the 4π configuration, the combined intensity exhibits a pattern resembling the radiation intensity distribution of a dipole emitter, with a Gaussian-shaped central lobe. Furthermore, at high numerical apertures, the focal volume can be reduced to values below $(0.5\ \lambda)^3$.

In the case of incoherent beam combining, the focal spot area also remains unchanged between the 4π and 2π configurations. Moreover, for systems with low numerical apertures, the depth of focus may extend over several centimeters.